\documentclass[a4paper,11pt]{article}
\pdfoutput=1 

\usepackage{jcappub} 

\usepackage[T1]{fontenc} 

\newcommand{\be}[1]{\begin{equation}\label{#1}}
\newcommand{\ee}{\end{equation}}
\newcommand{\ba}[1]{\begin{eqnarray}\label{#1}}
\newcommand{\ea}{\end{eqnarray}}
\newcommand{\rf}[1]{(\ref{#1})}

\title{\boldmath Suppression of matter density growth at scales exceeding\\ the cosmic screening length}

\author[a,1]{M. Eingorn,\note{Corresponding author.}}
\author[b,c,d]{E. Yilmaz,}
\author[d]{A.E. Y\"{u}kselci}
\author[b,c,e]{and A. Zhuk}

\affiliation[a]{Department of Mathematics and Physics, North Carolina Central University, \\1801 Fayetteville St., Durham, North Carolina 27707, U.S.A.}
\affiliation[b]{Center for Advanced Systems Understanding, Untermarkt 20, 02826 G\"{o}rlitz, Germany}
\affiliation[c]{Helmholtz-Zentrum Dresden-Rossendorf, Bautzner Landstra\ss e 400,\\ 01328 Dresden, Germany}
\affiliation[d]{Department of Physics, Istanbul Technical University, 34469 Maslak, Istanbul, Turkey}
\affiliation[e]{Astronomical Observatory, Odessa I.I. Mechnikov National University, \\ Dvoryanskaya St. 2, Odessa 65082, Ukraine}

\emailAdd{maxim.eingorn@gmail.com}
\emailAdd{e.yilmaz@hzdr.de}
\emailAdd{yukselcia@itu.edu.tr}
\emailAdd{ai.zhuk2@gmail.com}

\abstract{One of the main objectives of modern cosmology is to explain the origin and evolution of cosmic structures at different scales. The principal force responsible for the formation of such structures is gravity. In a general relativistic framework, we have shown that matter density contrasts do not grow over time at scales exceeding the cosmic screening length, which corresponds to a cosmological scale of the order of two to three gigaparsecs at the present time, at which gravitational interactions exhibit an exponential cut-off. This is a purely relativistic effect.
To demonstrate the suppression of density growth, we have performed N-body simulations in a box with a comoving size of $5.632\,{\rm Gpc}/h$ and  obtained the power spectrum of the mass density contrast. We have shown that it becomes independent of time for scales beyond the  screening length as a clear manifestation of the cosmic screening effect.}

\begin{document}
\maketitle
\flushbottom

\section{Introduction}
\label{sec:intro}

One of the most beautiful sights on Earth is the starry sky above our heads. From the creation of cosmogonic myths to recent studies in modern cosmology, humankind has sought to understand the origin of this natural phenomenon throughout history. Now, armed with telescopes operating at ever increasing precision, both on Earth and in space, we know that there exist various distinct structures ranging from planets and stars to galaxies that host them, and to groups and clusters of galaxies built up hierarchically. At cosmological scales, these structures form interconnected filaments with vast voids between them \cite{SDSS}. This is the so-called cosmic web. 
How has this cosmic web come into being, and how has it evolved into its current form that we observe? These are the questions that address some of the most significant challenges of modern cosmology.

There are different approaches to solving this problem. Analytical methods \cite{Bardeen,Mukhanov,Rubakov} are useful for describing the linear regime in which the matter density contrast remains small. 
This is relevant to the early stages of the evolution of the Universe as well as large-enough cosmological scales of the late Universe.
Unlike analytical methods, numerical simulations allow us to consider regions with both small and large density contrasts.
In Newtonian N-body simulations such as \cite{Gadget4}, interactions between massive particles are described by the Newtonian potential. These simulations may provide very accurate predictions for the $\Lambda$CDM model as long as large-scale corrections are taken into account, which also include the effect of relativistic components, and a correspondence with general relativity is to be worked out via particular \textit{dictionaries} \cite{a,Chisari,e}. Meanwhile, it is quite reasonable to expect that relativistic effects would already play a role in the equation for the gravitational potential. Indeed, it has been shown in a general relativistic framework in \cite{screening1,screening2,screening3,EinRus2} that the gravitational potential is described by an equation of Helmholtz-type and not of Poisson-type as in the Newtonian theory. The corresponding \textit{screening effect} implies that at scales larger than the screening length (which follows from the Helmholtz equation) the gravitational interaction is subject to an exponential cut-off and thus the growth of structures at such scales should be suppressed. In other words, the mass density contrast should no longer increase with time over larger domains. It is worth noting that the recent candidate for the largest-yet-observed structure in the Universe -- the Hercules-Corona Borealis Great Wall -- has dimensions of 2$\div 3 \,{\rm Gpc}$ \cite{32}, which coincide with this cut-off scale at the present time. 
 
It is of great interest to confirm  the above mentioned relativistic effect both analytically and experimentally. Suppression of the growth of the mass density contrast at large scales has previously been shown analytically in the hydrodynamic approximation, where peculiar velocities are described using the velocity potential \cite{duel,Ruslan1}. 
Modern observational catalogs, such as the SDSS \cite{SDSS}, do not cover regions that are of the same order of magnitude as the screening length in size, or greater; they cannot be used at the moment to test the desired effect. Yet, going beyond the hydrodynamic approximation, it is possible to perform numerical simulations based on Einstein equations, linearized with respect to metric corrections. This approach is the next level of verification of the screening effect. Such modeling, of course, should be carried out at scales that are much larger than the screening length at the present time. 

The remainder of the paper is organized as follows. In Section 2, we briefly present the weak-field expansion scheme of the cosmic screening approach and the equation for the gravitational potential. We obtain the matter power spectrum in Section 3, using a relativistic N-body simulation, and reveal the suppression in the density growth at various redshifts beyond the respective cut-off scales. The following Section 4 provides a comparison of the simulated power spectra with hydrodynamic approximation, which agree over a large range of modes, as long as the growth of inhomogeneities can be accurately described via the latter. We briefly summarize the results in Section 5.  

\section{Recap on the cosmic screening approach}
\label{sec:sec1}

Within the framework of general relativity, since the gravitational field is weak at all scales except near black holes and neutron stars, the space-time metric may conveniently be split into a homogeneous isotropic background and perturbations caused by inhomogeneities. Then, in the conformal Newtonian gauge, the perturbed Friedmann-Lema\^{\i}tre-Robertson-Walker metric \cite{Mukhanov,Rubakov,Durrer} for a spatially flat model reads
\ba{1.4}
ds^2 = a^2(\eta) \left[ \left(1+2\Phi \right)d\eta^2 - \left(1-2\Psi \right) \,\delta_{\alpha\beta}dx^{\alpha}dx^{\beta} \right], 
\ea
where the scale factor $a$ of the background depends only on the conformal time $\eta$, and $x^{\alpha}$ are comoving coordinates. Here, $\Phi$ and $\Psi$ are the scalar perturbations and $\Phi$ describes the gravitational potential. In the standard Lambda cold dark matter ($\Lambda$CDM) cosmology, in the absence of relativistic peculiar velocities, the first-order perturbations $\Phi$ and $\Psi$ are identical: $\Phi=\Psi$ (the model including relativistic species such as neutrinos has recently been considered, for instance, in \cite{MME}).

Recently, based on the metric in \rf{1.4} as well as the effective description of peculiar velocities, the equation for the gravitational potential in the cosmic screening approach was shown to take the form \cite{duel}
\be{2}
\bigtriangleup \Phi -\frac{a^2}{\lambda_{\rm{eff}}^2}\Phi =\frac{\kappa c^2}{2a}\delta\rho\, 
\ee
for the $\Lambda$CDM model (with $\Phi=\Psi$). Here, $\kappa \equiv 8\pi G_{\rm{N}}/c^4$ and $\delta\rho =\rho -\bar\rho$ is the mass density fluctuation; $\lambda_{\rm{eff}}(\eta)$ 
determines the characteristic length of the Yukawa cut-off for gravitational interactions, and  
relying on the cosmological parameters reported in \cite{planck2018} (Planck 2018), $\lambda_{{\rm{eff}},0}=2.57\,{\rm Gpc}$ at the present  time.

For a system of nonrelativistic point masses $m_n$ with the respective comoving radius-vectors ${\bf{r}}_n(\eta)$, the mass density reads $\rho =\sum_n\rho_n=\sum_n m_n \delta_{\rm{D}}({\bf r} -{\bf r}_n)$ and $\bar \rho$ corresponds to its average value. Given the definition of the corresponding energy-momentum tensor, $T^{ik}=\sum_n(m_n c^2/\sqrt{-g})(dx^{i}_n/d\eta)(dx^{k}_n/d\eta)(d\eta/ds_n)\delta_{\rm{D}}({\bf r} -{\bf r}_n)\,$, $i,k=0,1,2,3$ \cite{Landau}, where $g$ denotes the determinant of the metric, the time-time component of the linearized field equations and the longitudinal part of the time-space component read \cite{screening1}

\be{extra11}
\triangle\Phi-3\mathcal{H}(\Phi'+\mathcal{H}\Phi)=\frac{1}{2}\kappa a^2\left(\frac{c^2}{a^3}\delta\rho+\frac{3\overline{\rho} c^2}{a^3}\Phi\right)\,,
\ee

\be{extra13}
\Phi'+\mathcal{H}\Phi = -\frac{\kappa c^2}{2a}\frac{1}{4\pi}\sum_n m_n\frac{({\bf r} -{\bf r}_n)\tilde{\bf v}_n}{|{\bf r} -{\bf r}_n|^3}\,,
\ee
respectively. Above and hereafter the prime denotes the derivative with respect to conformal time, $\tilde{v}^\alpha_n\equiv dx^\alpha_n/d\eta$, and $\mathcal{H}\equiv a'/a$ is the conformal Hubble parameter. The term $\propto \overline{\rho}\Phi$ follows from the replacement $\rho\Phi\to\overline{\rho}\Phi$, as $|\delta\rho|\gg|\delta\rho\Phi|$ at all scales \cite{Chisari}, even though the inhomogeneous mass density is handled non-perturbatively. We note that extended discussions on the weak-field expansion scheme of this approach and perturbative ordering of source terms are provided in \cite{screening1,EinRus2}. Substituting \rf{extra13} into \rf{extra11}, one obtains
\be{extra14}
\triangle\Phi-\frac{3\kappa\overline{\rho}c^2}{2a}\Phi=\frac{\kappa c^2}{2a}\delta\rho-\frac{3\kappa c^2\mathcal{H}}{2a}\left(\frac{1}{4\pi}\sum_n m_n\frac{({\bf r} -{\bf r}_n)\tilde{\bf v}_n}{|{\bf r} -{\bf r}_n|^3}\right)\,.
\ee
This Helmholtz-type equation for the scalar potential was used to define a cutoff-scale, i.e. a \textit{screening length} for the gravitational force, that is $\lambda\equiv\sqrt{2a^3/(3\kappa\overline{\rho}c^2})$ \cite{screening1}.

Based on the ansatz $\Phi=(D_1/a)\phi({\bf r})$, introduced in \cite{Hahn}, relativistic theory counterparts of \rf{extra11} and \rf{extra13} take the form
\be{extra15}
\triangle\Phi-3\mathcal{H}\frac{D'_1}{D_1}\Phi=\frac{1}{2}\kappa a^2\left(\frac{c^2}{a^3}\delta\rho+\frac{3\overline{\rho} c^2}{a^3}\Phi\right)=\frac{1}{2}\kappa a^2\delta\varepsilon\,,
\ee    
now with the conventional source term $\propto\delta\varepsilon$, the energy density fluctuation that is equivalent to the term in parentheses, and  
\be{extra16}
\frac{D'_1}{D_1}\Phi=-\frac{1}{2}\kappa a^2\overline{\varepsilon}\nu\,,\;\; \overline{\varepsilon}\equiv \frac{\overline{\rho}c^2}{a^3}\,,
\ee
respectively. $D_1(\eta)$ is the linear growth factor, $\nu(\eta,\bf{r})$   stands for the velocity potential, and the comoving screening length implied by \rf{extra15} has the form $l\equiv1/\sqrt{3\mathcal{H}^2(d\ln D_1/d\ln a)}$ \cite{Hahn}.

It was argued in \cite{duel} that even though linear perturbation theory is applicable only to large-enough scales at late times, the expression in parentheses in \rf{extra14} can still be replaced by $\overline{\rho}\nu$, because below the scale of nonlinearity, where the theory breaks down, peculiar motion becomes negligible
and so does the last term of \rf{extra14}. Following this substitution, the novel Helmholtz equation reads
\be{extra17}
\triangle\Phi-\frac{3\kappa\overline{\rho}c^2}{2a}\Phi=\frac{\kappa c^2}{2a}\delta\rho-\frac{3\kappa \overline{\rho}c^2\mathcal{H}}{2a}\nu\,,
\ee   
and it still addresses all scales. Combining \rf{extra16} with the above equation, one eventually obtains \rf{2}, where $\lambda^{-2}_{\rm eff}\equiv\lambda^{-2}+(al)^{-2}$.

\section{Power spectrum of the matter density contrast}
   
The distribution of the mass density contrast $\delta({\bf{r}},z) = \delta\rho({\bf{r}},z)/\bar\rho$  can be described via the power spectrum \cite{Durrer,gevNat}:
\be{3}
4\pi k^3\langle \widehat\delta({\bf{k}},z)\, \widehat\delta^*({\bf{k}}',z)\rangle =(2\pi)^3\delta_{\rm D} ({\bf k}-{\bf k}')P_{\delta}(k,z)\, ,
\ee
where $\widehat\delta({\bf{k}},z)$ is the Fourier transform of $\delta({\bf{r}},z)$, and $\widehat\delta^*({\bf{k}},z)$ is its complex conjugate. Another definition of the power spectrum $\tilde P_{\delta}(k,z)$ is also often used in the literature \cite{Rubakov}, which is related to the above-mentioned one as follows: $\tilde P_{\delta}(k,z)=(2\pi^2/k^3)P_{\delta}(k,z)$. The square root of $\tilde P_{\delta}(k,z)$ specifies a characteristic value of the density contrast at the scale $l_c\sim a/k$ at redshift $z$. 

To calculate $P_{\delta}(k,z)$ in the $\Lambda$CDM model, we resort to relativistic N-body simulations. The basis for these are Einstein equations, in which matter is represented by massive discrete particles, playing the role of galaxies or groups of galaxies etc. Such equations are presented, for example, in \cite{gevNat} as used in the relativistic code \textit{gevolution} and in the previous section for the cosmic screening approach. In the case of nonrelativistic particles, the difference between the two is that equations in \textit{gevolution} include not only linear terms, but also those which are quadratic in scalar perturbations. As a result, metric corrections represent a mixture of first- and second-order quantities. The \textit{gevolution} expansion also includes a term $\propto \Phi^{\prime}$, which, at first glance, does not appear in \rf{2}, but this contribution is accounted-for in deriving $\lambda_{\rm eff}$. The code \cite{gevNat,gevolution} allows to calculate the power spectra of metric perturbations as well as that of the energy density contrast, and can easily be modified to study the behaviour of these quantities in the cosmic screening approach. Indeed, a comparative analysis was carried out in \cite{scr-gev}, and it was shown that despite the presence of second-order contributions to metric perturbations in \textit{gevolution}, the power spectra of these potentials demonstrate very good agreement between the two schemes. Hence, the effect of second-order admixtures is small, as it should be, and the simpler form equations in the screening approach helps save almost 40\% of CPU hours. Now, to obtain $P_{\delta}(k,z)$, we have modified \textit{gevolution} \cite{gevNat,gevolution} and replaced its default set of equations by that of the screening approach. This means that to single out the term $\delta\rho =\rho -\bar\rho$ on the right hand side of the field equation (2.9) in \cite{gevolution}, we have moved the additional $\Phi$-dependent term in the perturbed energy-momentum tensor element $\delta\varepsilon =\delta T_{0}^0=T_{0}^0-\overline{T_{0}^0}$ (where $T_{0}^0$ is given by  (3.7) of \cite{gevolution}) to the left hand side (see \cite{scr-gev}). We have also dropped all  second-order corrections and demanded that all particles are nonrelativistic. Eventually, particles' momenta ${\bf q}_n$ and   the gravitational potential $\Phi$ were evolved according to
\be{extra1}
\frac{dq_n^{\alpha}}{d\eta}=-m_n a\frac{\partial \Phi}{\partial x^\alpha}\,
\ee
and
\be{extra2}
\triangle\Phi-3\mathcal{H}\Phi'-3\Phi\left(\mathcal{H}^2+4\pi G_{\rm{N}}\overline{\rho}/a\right)
=4\pi G_{\rm{N}} a^2\left(\sum_n m_n\delta_{\rm D}({\bf r}-{\bf r}_n)/a^3-\overline{T^0_0}\right)\,,
\ee
respectively (for $c=1$). We find it useful to stress once again that the matter power spectrum obtained in the screening approach is that of the mass density (contrast) $\delta\rho/\overline{\rho}$ and \textit{not} the energy density (contrast) $\delta\varepsilon/\overline{\varepsilon}$. A comparison between the two has been studied recently in \cite{massden} and it has been shown that the power spectra curves for these two quantities deviate from one another significantly at large scales and for the redshifts investigated here.

\begin{figure}[h!]
\centering 
\includegraphics[width=.85\textwidth]{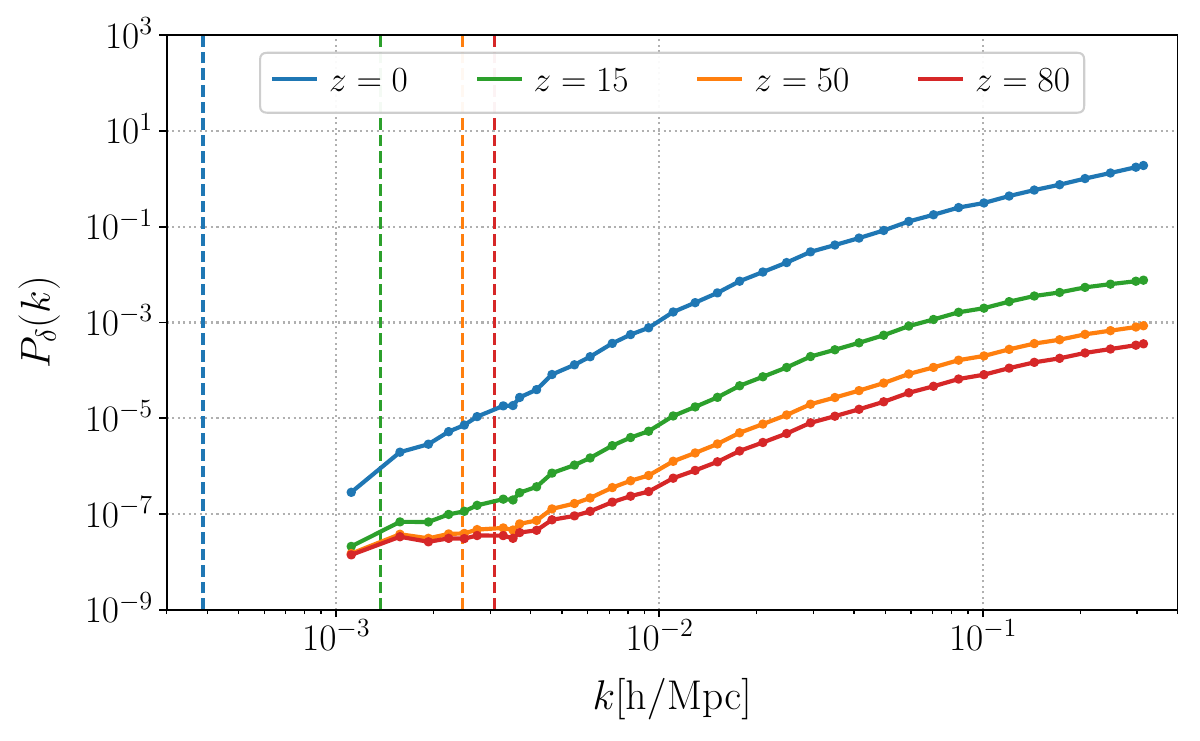}
\caption{\label{fig:1} {\bf{Power spectrum $P_{\delta}(k,z)$ of the mass density contrast.}} We present the power spectra for redshifts $z=0$ (blue curve), $z=15$ (green curve), $z=50$ (orange curve) and $z=80$ (red curve) from top to bottom.	
 The box size was set to $5632\,{\rm Mpc}/h$ for this simulation. Dashed vertical lines indicate the comoving momenta  $k_{\rm scr}(z) =a/\lambda_{\rm eff}(z)= 1/[(1+z)\lambda_{\rm eff}(z)]$ for the respective screening lengths. In the area to the left of the dashed vertical orange line, the orange and red curves converge towards each other. This indicates the suppression of growth of the density contrast with time at scales beyond the screening length.}
\end{figure}

\begin{figure}[h!]
\centering 
\includegraphics[width=.85\textwidth]{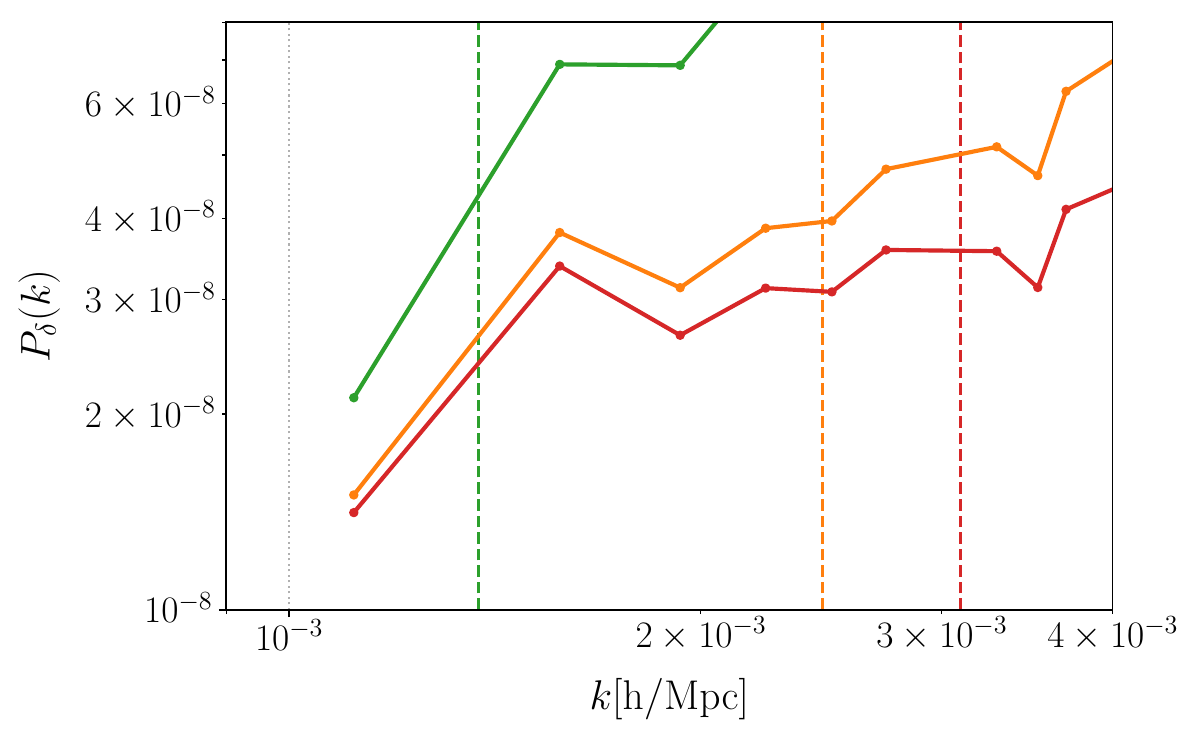}
\caption{\label{fig:2} {\bf{Behaviour of $P_{\delta}(k,z)$ in the vicinity of screening lengths.}} From top to bottom, green, orange and red curves denote the power spectra at $z=15,\, 50,\, 80$, respectively.}
\end{figure}

In order to observe the suppression of the growth of density contrasts beyond the screening length, it is necessary to run the simulations in boxes larger than its current value. We have therefore used a box with a comoving size of 5.632 ${\rm Gpc}/h$ ($h$ is the dimensionless Hubble constant), which coincides with the physical size of the box at the present time, and is greater than $\lambda_{{\rm{eff}},0}$. The resolution was set to $2\, {\rm Mpc}/h$ with $\approx 22\times10^9$ particles. We have adopted the conventional normalization for the scale factor, i.e. $a_0=1$, and used the cosmological parameters $H_0 = 67.66 \,  {\rm km}\, {\rm s}^{-1} {\rm Mpc}^{-1}$, $\Omega_{\rm M}h^2 = \Omega_{\rm b}h^2+\Omega_{\rm c}h^2=0.14175$, where $h=0.6766$,
and $\Omega_{\Lambda}=1-\Omega_{\rm M}$ \cite{planck2018}. The simulation was run assuming no distinction between baryons and cold dark matter. The resulting power spectra $P_{\delta}(k,z)$ of the matter density contrast at redshifts $z = 0,\,15,\,50,\,80$ are presented in figures \ref{fig:1} and \ref{fig:2}, where the latter zooms into the region in which the $z=50$ and $z=80$ curves converge towards each other. Dashed vertical lines in both figures correspond to screening lengths in momentum space: $k_{\rm scr}(z) =a/\lambda_{\rm eff}(z)$. The two curves in these graphs, describing the power spectra at $z=80$ (red) and $z=50$ (orange), reach well beyond their respective screening lengths. The area to the left of the orange vertical line is greater than the screening lengths for both curves. The red and orange curves merge in this area, which means that density contrasts no longer grow with time (with $z$) in this region. This is a clear manifestation of the screening effect: we see that the redshift-dependent growth of matter density contrast is suppressed at scales exceeding the effective screening length \cite{duel} (i.e. for the momenta $k<a/\lambda_{\rm eff}=[(1+z)\lambda_{\rm eff}]^{-1}$),
\be{5}
\lambda_{\rm{eff}}(z)=\left(\frac{c^2a^2H}{3}\int_0^a\frac{d\tilde a}{\tilde a^3 H^3}\right)^{1/2}\, ,
\ee
where the Hubble parameter $H=H_0\sqrt{\Omega_{\rm M}(a_0/a)^3 +\Omega_{\Lambda}} $. A weak growth of density contrast begins to take place already between the vertical orange and red lines, and to the right of the red vertical line, i.e. at scales smaller than the screening lengths for both curves, we observe that the regular growth pattern is restored.  


\section{Cosmic screening in hydrodynamic approximation}
\begin{figure}[tbp]
\centering 
\includegraphics[width=.85\textwidth]{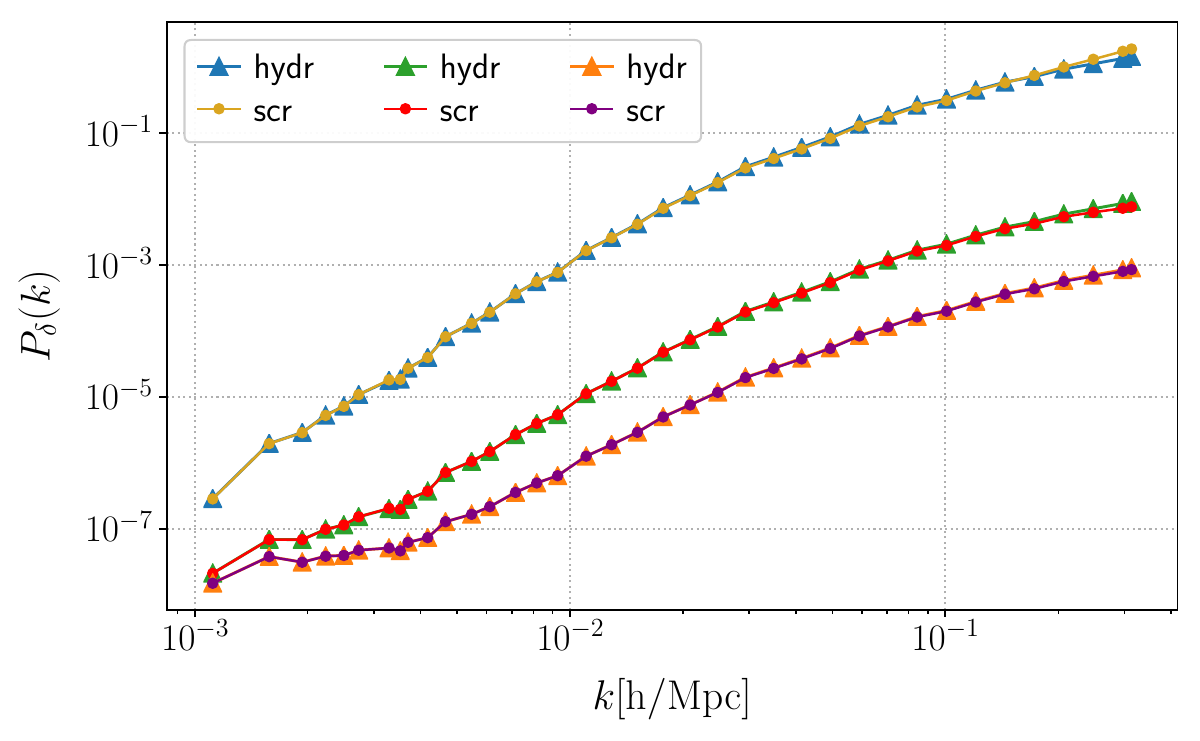}
\caption{\label{fig:3} {\bf{Comparison of the power spectra obtained in the N-body simulation and hydrodynamic approximation.}} From top to bottom, gold ($z=0$), red ($z=15$) and purple ($z=50$) curves correspond to the N-body simulation results. Blue ($z=0$), green ($z=15$) and orange ($z=50$) curves, again, from top to bottom, are obtained using eq.~\rf{4} and well approximate the curves obtained in the simulation. Deviation becomes manifest in the region of large momenta, where the hydrodynamic approximation breaks down.}
\end{figure}

Now, we turn to the power spectrum of the mass density contrast in hydrodynamic approximation valid for large scales (small $k$). Within the cosmic screening approach, the Fourier transform of the matter density fluctuation at large cosmological scales reads \cite{duel}
\be{6}
\widehat{\delta\rho} ({\bf k},z)\propto\widehat\phi ({\bf k})\left(D_1k^2+3\mathcal{H}D_1'+\frac{3\kappa\bar\rho c^2 }{2a}D_1\right)\, ,
\ee
where, without loss of generality, the growing mode of the growth factor $D_1(z)$ may be presented as $\left(\mathcal H/a\right)\int_0^a d{\tilde a}/{\mathcal H}^3$ \cite{D1}, while $\widehat\phi ({\bf k})$ is a function of ${\bf k}$ only. Taking into account the definition \rf{3} and the expression \rf{6} as well as the equality $3\mathcal{H}^2-3\mathcal{H}'=3\kappa\bar\rho c^2/(2a)$ (which follows directly from the Friedmann equations), the formula for the matter power spectrum may be expressed as
\be{8}
P_{\delta}(k,z)= F(k) \left[D_1k^2+3\mathcal{H}D_1'+(3\mathcal{H}^2-3\mathcal{H}')D_1\right]^2\,\ee
for $\Lambda$CDM cosmology. Employing the explicit expression for $D_1$, the above expression takes the more compact form 
\be{4}
P_\delta(k,z)=F(k)\left[D_1(z)k^2+3\right]^2\, .
\ee
An important point worth highlighting is that the prefactor $F(k)$ has no dependence on time (equivalently, on redshift $z$);  $z$-dependence is completely defined by the expression in brackets, and is easily calculated. 
Therefore, knowing how $F$ depends on $k$, one may obtain the power spectrum for any $z$ using the formula \rf{4}. This dependence can be revealed using one of the curves shown in figure~\ref{fig:1}. In our case, for this purpose, we used the one for $z=80$: $F(k)=P_\delta(k,80)/\left[D_1(80)k^2+3\right]^2$. We have constructed the power spectra at $z =0,\, 15,\,50$, as shown in figure~\ref{fig:3}. Here, the curves obtained in the case of small density contrasts provide a good approximation for the corresponding power spectra, which follow from the N-body simulation. The deviation takes place only at large $k$ (small distances), where large density contrasts play a significant role and the hydrodynamic approximation fails to accurately describe the growth of inhomogeneities.

At $z\gg 1$, the Universe is going through the matter-dominated era for which the conformal Hubble parameter ${\mathcal H}=\kappa \bar\rho c^2/(3a)$ and, consequently, $D_1=a$; at the same time, $\lambda_{\rm{eff}}=\sqrt{2a^3/(5\kappa \bar\rho c^2)}$ \cite{duel}. During this evolution stage we get
\be{7}
\widehat{\delta\rho} \propto k^2\widehat\phi\left(a +\frac{5\kappa\bar\rho c^2}{2}\frac{1}{k^2}\right) = k^2\widehat \phi \left(a+\frac{a^3}{\lambda_{\rm{eff}}^2}\frac{1}{k^2}\right)=k^2\widehat \phi\left(a+a\frac{k_{\rm{scr}}^2}{k^2}\right)\, ,
\ee
where $k_{\rm{scr}}=a/\lambda_{\rm{eff}}\propto1/\sqrt{a}$ in the matter-dominated era, so the second term in brackets does not depend on $z$. Therefore, for $k \ll k_{\rm{scr}}(z)$, mass density fluctuations do not depend on the redshift. This is an obvious manifestation of cosmic screening. It also gives grounds for drawing the vertical lines in figure~\ref{fig:1}.

Returning to eq.~\rf{7}, in the opposite limit $k\gg k_{\rm{scr}}(z)$, the ratio $\widehat{\delta\rho}({\bf k},z_1)/\widehat{\delta\rho}({\bf k},z_2)\approx (1+z_2)/(1+z_1)\equiv R_{z_1,z_2}$ does not depend on momentum ${\bf k}$. Therefore, for this range of $k$, power spectrum curves corresponding to mass density contrasts should be parallel to one another, as also seen in figure~\ref{fig:1}. Moreover, it is possible to estimate the difference between the curves easily. For instance, $R_{50,80}\approx 1.59$, and it is quite close to the value $\sqrt{P_{\delta}(z=50)/P_{\delta}(z=80)}\approx 1.57$ obtained from the simulation outputs, that are $P_{\delta}(z=50)\approx19.5\times 10^{-5}$ and $P_{\delta}(z=80)\approx7.92\times 10^{-5}$ for the sample value of $k=10^{-1}$.

\section{Conclusion}

In this paper we have presented the matter power spectrum for $\Lambda$CDM cosmology and revealed that the redshift-dependent growth of density contrasts becomes suppressed at scales larger than the characteristic cut-off scale introduced in the general relativistic screening approach. An explicit demonstration of the power spectra for four different redshifts has been provided via an N-body simulation, carried out in a comoving box of $5.632\, {\rm Gpc}/h$, which well exceeds the current value of the screening length, that is $\lambda_{\rm eff,0}=2.57\, {\rm Gpc}$. It is important to emphasize that screening of gravity at large cosmological scales is a purely relativistic effect. The energy-momentum tensor, which is the source for the gravitational field, itself depends on the gravitational field (on metric coefficients).  Because of that, the equation for the gravitational potential acquires a term, which is proportional again to the gravitational potential. This screening mechanism intrinsic to the Helmholtz equation implies that there exists a time-dependent upper bound for the dimensions of individual cosmic structures as confirmed by the behaviour of the matter power spectrum obtained in the present work.  
 
Then, we have proceeded with the formulation of the power spectrum of the mass density contrast in hydrodynamic approximation. A comparison with the simulation outcomes has been performed and it has been shown that the approximation remains valid up to large modes, as it should, until nonlinear growth of density contrasts begins to take place at small scales.

\acknowledgments

This work was partially supported by the Center for Advanced Systems Understanding (CASUS) which is financed by Germany's Federal Ministry of Education and Research (BMBF) and by the Saxon state government out of the State budget approved by the Saxon State Parliament. Computing resources used in this work were provided by the National Center for High Performance Computing of T\"{u}rkiye (UHeM) under grant number 4007162019.

\ 

{\bf\large\noindent Data availability}

\ 

\noindent The datasets generated in this work are publicly available in the Rossendorf Data Repository (RODARE) and may be accessed via the link \newblock\href{https://rodare.hzdr.de/record/2371}{{https://rodare.hzdr.de/record/2371}}



\begin{thebibliography}{99}

\bibitem{SDSS} Abdurro'uf et al., \emph{The Seventeenth Data Release of the Sloan Digital Sky Surveys: Complete Release of MaNGA, MaStar, and APOGEE-2 Data}, \emph{ApJS} {\bf 259} (2022) 35.
\bibitem{Bardeen} 
J. Bardeen, \emph{Gauge-invariant cosmological perturbations}, \emph{Phys. Rev. D} {\bf 22} (1980) 1882.
\bibitem{Mukhanov} 
V.F. Mukhanov, \emph{Physical foundations of cosmology}, Cambridge University Press (2005).
\bibitem{Rubakov} 
D.S. Gorbunov and V.A. Rubakov, \emph{Introduction to the Theory of the Early Universe: Cosmological Perturbations and Inflationary Theory},
World Scientific Publishing (2011).
\bibitem{Gadget4} V. Springel, R. Pakmor, O. Zier and M. Reinecke, \emph{Simulating cosmic structure formation with the GADGET-4 code}, \emph{Mon. Not. R. Astron. Soc.} {\bf 506} (2021) 2871. 
\bibitem{a}  S.R. Green and R.M. Wald, \emph{Newtonian and Relativistic Cosmologies}, \emph{Phys. Rev. D} {\bf 85} (2012) 063512.
\bibitem{Chisari}N.E. Chisari and M. Zaldarriaga, \emph{Connection between Newtonian simulations and general relativity}, \emph{Phys. Rev. D} {\bf 83} (2011) 123505.
\bibitem{e} C. Fidler, T. Tram, C. Rampf,  R. Crittenden, K. Koyama and D. Wands, \emph{General relativistic weak-field limit and Newtonian N-body simulations}, \emph{JCAP} {\bf 012} (2017) 022. 
\bibitem{screening1}
M. Eingorn, \emph{First-order cosmological perturbations engendered by point-like masses}, \emph{Astrophys. J.} {\bf 825} (2016) 84.
\bibitem{screening2}
M. Eingorn, C. Kiefer, and A. Zhuk, \emph{Scalar and vector perturbations in a universe with discrete and continuous matter sources}, \emph{JCAP} {\bf 09} (2016) 032.
\bibitem{screening3}
M. Eingorn, C. Kiefer, and A. Zhuk, \emph{Cosmic screening of the gravitational interaction}, \emph{Int. J. Mod. Phys. D} {\bf 26} (2017) 1743012.
\bibitem{EinRus2}	
R. Brilenkov and M. Eingorn, \emph{Second-order cosmological perturbations engendered by point-like masses}, \emph{Astrophys. J.} {\bf 845} (2017) 153.
\bibitem{32}
I. Horvath, Z. Bagoly, J. Hakkila, and V. Toth,  \emph{New data support the existence of the Hercules-Corona Borealis Great Wall}, \emph{Astron. Astrophys.} {\bf 584} (2015) A48.
\bibitem{duel}
E. Canay and M. Eingorn, \emph{Duel of cosmological screening lengths}, \emph{Phys. Dark
Univ.} {\bf 29} (2020) 100565.
\bibitem{Ruslan1}
M. Eingorn and R. Brilenkov,  \emph{Perfect fluids with $\omega = \mathrm{const}$ as sources of scalar cosmological perturbations}, \emph{Phys. Dark Univ.} {\bf 17} (2017) 63.
\bibitem{Durrer} 
R. Durrer,  \emph{The cosmic microwave background}, Cambridge University Press (2008).
\bibitem{MME}	
M. Brilenkov, E. Canay, and M. Eingorn,  \emph{Cosmological perturbations engendered by discrete relativistic species}, \emph{Eur. Phys. J. C} {\bf 83} (2023) 601.
\bibitem{planck2018} 
N. Aghanim, et al. [Planck Collaboration], \emph{Planck 2018 results. VI. Cosmological parameters}, \emph{Astron. Astrophys.} {\bf 641} (2020) A6.
\bibitem{Landau}
L.D. Landau and E.M. Lifshitz, \emph{Course of Theoretical Physics Series, Vol. 2, The Classical Theory of Fields}. Oxford Pergamon Press (2000).
\bibitem{Hahn}
O. Hahn and A. Paranjape, \emph{General relativistic 'screening' in cosmological simulations}, \emph{Phys. Rev. D} {\bf 94} (2016) 083511.
\bibitem{gevNat}
J. Adamek, D. Daverio, R. Durrer, and M. Kunz,   \emph{General relativity and cosmic structure formation}, \emph{Nature Phys.} {\bf 12} (2016) 346.
\bibitem{gevolution}	
J. Adamek, D. Daverio, R. Durrer and M. Kunz,   \emph{gevolution: a cosmological N-body code based on General Relativity}, \emph{JCAP} {\bf 07} (2016) 053.
\bibitem{scr-gev} 
M. Eingorn, A.E. Y\"{u}kselci and A. Zhuk,  \emph{Screening vs. gevolution: in chase of a perfect cosmological simulation code}, \emph{Phys. Lett. B} {\bf 826} (2022) 136911.
\bibitem{massden} 
M. Eingorn, E. Yilmaz, A.E. Y\"{u}kselci and A. Zhuk,  \emph{Mass density vs. energy density at cosmological scales}, \emph{Phys. Lett. B} {\bf 851} (2024) 138564.
 \bibitem{D1}
F. Bernardeau, S. Colombi,  E. Gaztanaga, and  R. Scoccimarro, \emph{Large-scale structure of the Universe and cosmological perturbation theory}, \emph{Phys. Rept.} {\bf 367} (2002) 1.






\end{thebibliography}
\end{document}